\documentclass[journal,transmag]{IEEEtran}
\usepackage{hyperref,color,breqn,multirow}
\usepackage[top=0.7in, left=0.65in]{geometry}
\usepackage{graphicx}
\usepackage{hyperref}
\begin{document}
\title{Design of a loop-gap resonator with bimodal uniform fields using finite element analysis}
\author{\IEEEauthorblockN{Matthew M. Libersky\IEEEauthorrefmark{1}\thanks{Work at Caltech supported by the U.S. Department of Energy Office of Basic Energy Sciences, grant DE-SC0014866.},
Daniel M. Silevitch\IEEEauthorrefmark{1},
Ammar Kouki\IEEEauthorrefmark{2}}
\IEEEauthorblockA{\IEEEauthorrefmark{1}Division of Physics, Math, and Astronomy, California Institute of Technology, Pasadena, CA 91125 USA}
\IEEEauthorblockA{\IEEEauthorrefmark{2}\'{E}cole de technologie sup\'{e}rieure, Montreal, QC, Canada}}

\IEEEtitleabstractindextext{%
\begin{abstract}
The loop-gap resonator (LGR) was originally developed to provide a uniform microwave magnetic field on a sample for electron spin resonance (ESR) experiments. The LGR is composed of one or more loops and gaps acting as inductances and capacitances respectively. Typical LGR designs produce a uniform field on a sample at a single resonant frequency, but for certain experiments it is necessary to study the response of a material to uniform fields at multiple frequencies applied simultaneously. In this work we develop an empirical design procedure using finite element method calculations to design an asymmetric loop-gap resonator with uniform fields at two frequencies in the same sample volume and analyze the field uniformity, frequency tunability and filling factors, providing comparison to a manufactured device. 
\end{abstract}

}
 \IEEEoverridecommandlockouts
\IEEEpubid{\makebox[\columnwidth]{978-1-5386-6425-4/19/\$31.00~
		\copyright2019
		IEEE \hfill}
	 \hspace{\columnsep}\makebox[\columnwidth]{ }} 
\maketitle

\section{Introduction}
\IEEEPARstart{O}{riginally developed} for electron spin resonance (ESR) experiments in the S-band, loop-gap resonators (LGRs) have proven useful in a wide variety of experiments due to their ability to produce a uniform microwave magnetic field concentrated on a sample volume~\cite{meh}. In contrast to cavity and transmission line resonators where the physical dimensions are tied to the wavelength at the resonant frequency, LGRs are effective lumped element devices with separate inductive and capacitive elements, allowing for smaller mode volumes that are more practical for confined geometries and spectroscopy of small samples. \let\thefootnote\relax\footnote{\copyright2019 IEEE. Personal use of this material is permitted. Permission from IEEE must be obtained for all other uses, in any current or future media, including reprinting/republishing this material for advertising or promotional purposes, creating new collective works, for resale or redistribution to servers or lists, or reuse of any copyrighted component of this work in other works.}

The simplest LGR design consists of a metal tube with a narrow slot along its axis~\cite{meh,opie}. The inductance of the structure is determined by the loop, while the gap contributes a capacitance. To a good approximation, the field inside the loop is purely magnetic, with the electric field confined to the gap. A refinement of this approach~\cite{3l2g} uses three loops connected by two gaps to provide a controlled return flux path and reduce the strong radiation loss associated with the single-loop resonator's dipole field pattern. 

Due to the flexibility in design, field uniformity, and optical access through the sample, resonators using this design are useful for a wide range of experimental studies, such as transduction between microwave and optical states in quantum computing applications~\cite{ball}.

\section{Requirements and Motivation for Resonator Design}
When a sample with complex susceptibility $\chi = \chi'-j\chi''$ is inserted into a resonator, the reactive component $\chi'$ modifies the resonant frequency $f_0$ and the dispersive component $\chi''$ modifies the quality factor $Q$:

\begin{eqnarray}
\frac{\Delta f_0}{f_0} &= \frac{\chi'H^2 \Delta v}{2\int_{\mathrm{res}}H^2d\tau} \\
\Delta\left(\frac{1}{2Q}\right)  &= \frac{\chi''H^2 \Delta v}{2\int_{\mathrm{res}} H^2d\tau},
\end{eqnarray}
where $\Delta v$ is the volume of the sample and the integral is over the entire volume of the resonator~\cite{chen_book}. In ESR experiments this susceptibility has a characteristic peak at an ESR transition. When this frequency is off-resonance to all electronic level transitions, this quantity probes the static susceptibility of the sample, providing information about the bulk magnetic properties of a sample~\cite{chen_book}. By contrast, studying the on-resonant response of a sample at multiple frequencies can provide information on the spin Hamiltonian~\cite{multi}.

While multifrequency studies are typically performed by retuning a resonator and carrying out separate measurements, there are advantages to having a uniform field at two distinct frequencies simultaneously. For instance, many ESR experiments require temperatures of 0.1 K or below, and thus entails the use of a $^3$He/$^4$He dilution refrigerator. Cycling a dilution refrigerator to room temperature to retune a resonator can require hours or days, so the convenience of applying fields to a sample at multiple frequencies is considerable. More importantly, pump-probe spectroscopy with simultaneously-applied fields at two frequencies is a common technique \cite{multi}. However, little work has been done to develop a microwave resonator that can produce uniform, resonantly enhanced, ac magnetic fields at frequencies spaced by a few GHz. 

\begin{figure}[h]
\begin{center}
		\includegraphics[scale=0.8]{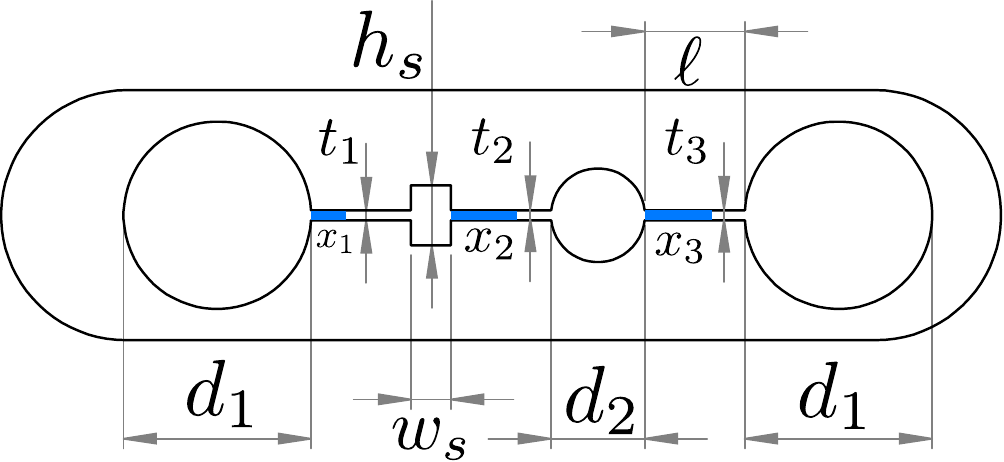}
\end{center}
	\caption{	\label{fig:dim}Illustration of the resonator design used in this work.}
\end{figure}

The resonator design studied in this work is shown in Fig.~\ref{fig:dim}. This design is a variation of a typical four-loop three-gap resonator, with asymmetry introduced to provide degrees of freedom to control the mode splitting. One loop is rectangular  to accommodate the rectangular single-crystal samples often used in ESR and quantum magnetism studies. The dimensions of the rectangular loop are fixed by the sample size (here we use 4x2 mm), and we choose to fix the gap lengths $\ell_i$ at 5~mm. We also make the the gaps a uniform width $t = 330~\mu\mathrm{m}$; the capacitance of the gaps can be independently tuned post-fabrication by introducing sapphire sheet of variable width fractions $x_i$ into the gaps.

In a three-loop two-gap design the outer loops serve as the return flux path for the center loop, so increasing the size of these loops enhances the flux density in the central loop. A similar principle applies here, so the outer loops are set as large as is practical. The size of the circular center loop plays a role in determining the relative filling factors of the two modes, as will be examined later.

The goals of our analysis are to measure the field uniformity for the rectangular loop while understanding the dependence of the mode spacing and filling factors on these parameters in order to guide design choices.

\section{COMPUTATIONAL ANALYSIS}
A 3D CAD model of the empty resonator is constructed using SOLIDWORKS~\cite{SW}. The resonator body is made of copper with the loop and gap structure cut out and surrounded by a few mm of air on either side. To compute the resonant frequencies and corresponding field modes, a generalized eigenmode analysis is carried out using an edge-based finite element formulation as implemented in EMWorks' high frequency simulator (HFWorks) \cite{EM}. This formulation ensures that no spurious modes are found and allows the use of the imperfect conductor boundary condition at the resonator's copper walls. To ensure higher precision of the resonant field distribution in the loop and gap regions, mesh refinement is used.  Mode power normalization is also employed to produce proper field level comparison plots.

\begin{figure}[h]
\begin{center}
		\includegraphics[scale=0.5]{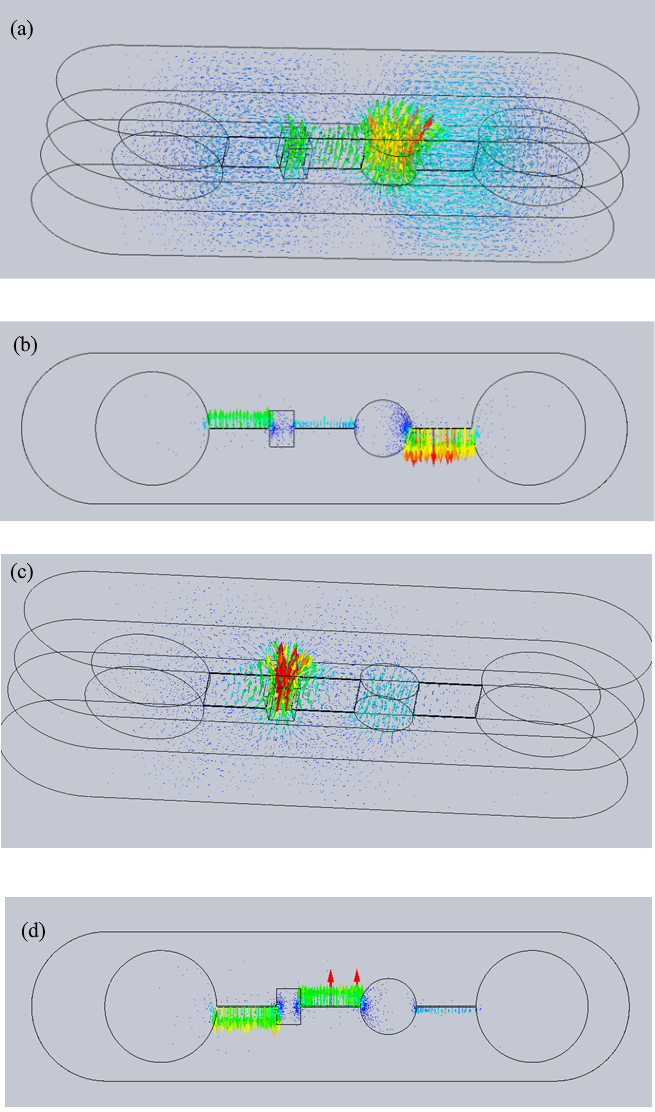}
\end{center}
\caption{\label{fig:vector}Calculated magnetic (a,c) and electric (b,d) fields of the low- (a,b) and high-frequency (c,d) modes.}
	
\end{figure}

Fig.~\ref{fig:vector} shows the magnetic and electric field distributions of the two desired modes. While both modes have flux through the sample loop, a mode-splitting occurs because the return flux paths for the two modes are different. In the low-frequency mode, the magnetic fields in the two center loops oscillate in-phase, using the two loops on the ends as return flux paths, whereas for the high-frequency mode the return flux path is predominantly the second center loop. 

\begin{figure}[h]
	\begin{center}
	\includegraphics[scale=0.7]{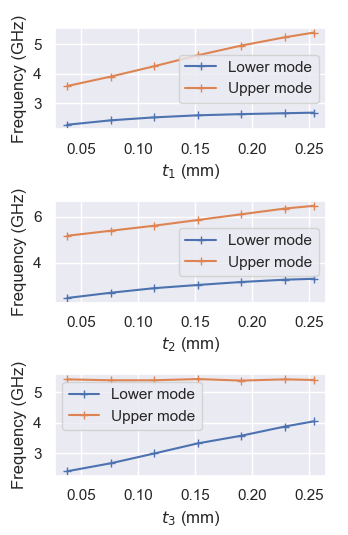}
	\end{center}
	\caption{	\label{fig:tune}
Calculated variation of the mode frequencies by tuning of the left, center, and right gap thicknesses. The low-frequency mode is largely insensitive to variation of the left gap while the high-frequency mode is independent of the right gap, allowing each mode to be tuned independently with dielectric loading post-fabrication.}
\end{figure}

The calculated electric field distribution is also qualitatively different between the two modes, with only the high-frequency mode having a significant electric field in the central gap. Fig.~\ref{fig:tune} shows the dependence of the mode frequencies on the three gap widths, with the low and high modes largely independently tunable using the right and left gaps respectively.

The filling factor, defined as 

\begin{equation}
\eta = \frac{\int_{\mathrm{sample}} H^2d\tau}{\int_{\mathrm{all}}H^2d\tau}, 
\end{equation}
is also an important parameter for sensitivity in spectroscopy and entering the regime of strong coupling between the resonator and a spin ensemble~\cite{angerer}. As mentioned previously, cavity and transmission line resonators have mode volumes tied intrinsically to the resonant wavelength, and thus have quite low filling factors for typical mm-size samples in the regime of a few GHz, making them impractical for spectroscopy in this range. 

\begin{figure}[h]
\begin{center}
		\includegraphics[scale=0.7]{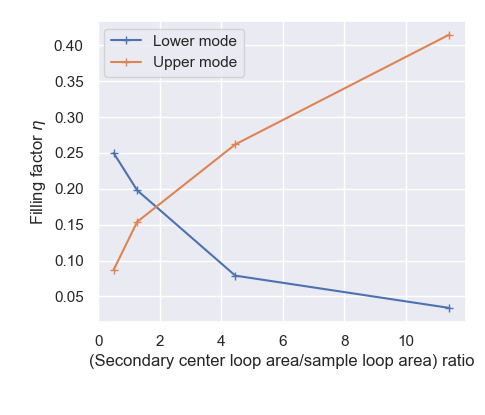}
\end{center}
	\caption{	\label{fig:fill}
Variation of the filling factors of the two modes as the size of the circular center loop is varied, as a function of the ratio between the area of the sample loop and the circular loop. A tradeoff is apparent due to the different return flux patterns of the two modes.}
\end{figure}

The filling factor is calculated here by numerically evaluating the integrals over a grid, with results shown in Fig.~\ref{fig:fill} for different sizes of the non-sample (circular) center loop. 

A tradeoff in the filling factors of the two modes is apparent, due to the different return flux patterns. The low mode essentially sees each central loop acting as the return path for flux through each adjacent outer loop, while for the high mode the sample loop predominantly serves as the return flux path for both the far-left loop and secondary center loop. This leads to a simple interpretation of these modes in terms of the three-loop two-gap structure discussed previously; in the lower mode both center loops the magnetic fields oscillate in phase and act as the single center loop in the three-loop two-gap geometry, while the higher mode corresponds to such a structure excluding one side loop.

\begin{figure}[h]
	\begin{center}
		\includegraphics[scale=0.65]{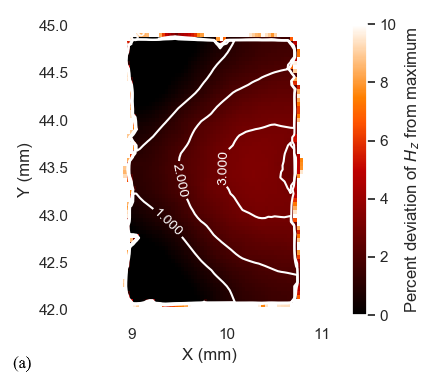}
		
		\includegraphics[scale=0.65]{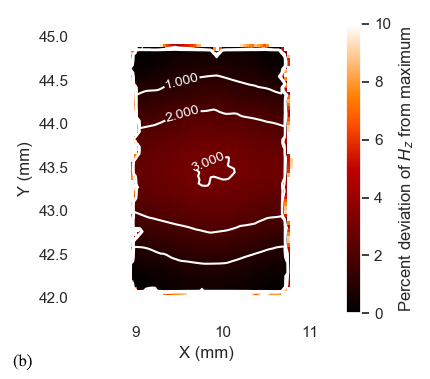}
	\end{center}
	\caption{Contour plots of $H_z$ within the sample volume for the low- (a) and high-frequency (b) modes. }
	\label{fig:contour}
\end{figure}

The field uniformity within the sample volume is also important for uniform saturation of a spin ensemble~\cite{angerer}. In Fig.~\ref{fig:contour} cuts through the midplane of the sample volume are shown. The maximum variation of the magnetic field component perpendicular to this plane is approximately 3\% for each mode, which is comparable to single-mode 3D lumped-element resonators~\cite{angerer}. 

\section{EXPERIMENTAL IMPLEMENTATION}
\begin{figure}[h]
\begin{center}
		\includegraphics[scale=0.45]{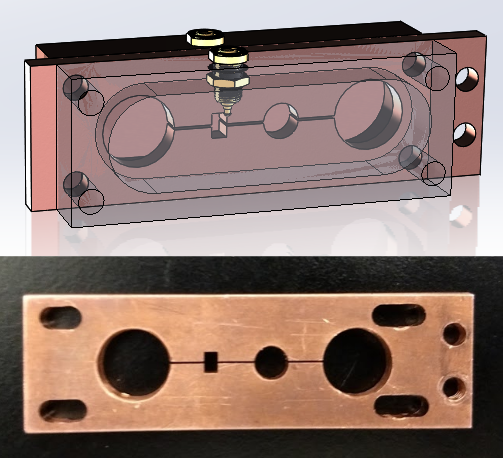}
\end{center}
	\caption{	\label{fig:exp} CAD model (top) of the resonator assembly including coupling pins and enclosure, and photograph (bottom) of the resonator manufactured using wire EDM.}
\end{figure}

\begin{figure}[h]
\begin{center}
		\includegraphics[scale=0.75]{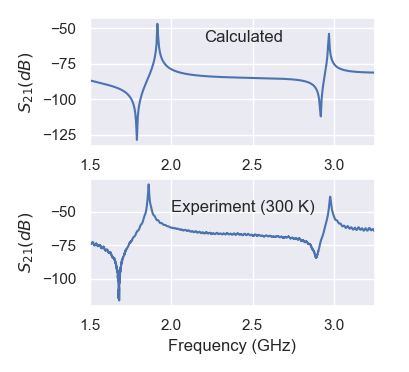}
\end{center}
	\caption{\label{fig:s21}Calculated (top) and experimental (bottom) transmission of this resonator design. }
\end{figure}

These structures are built using wire electrical discharge machining (EDM), allowing for fabrication of thin gaps without the losses introduced by bolting separate halves together \cite{3l2g}. A photograph of a resonator fabricated out of a single piece of oxygen-free copper using wire EDM is shown in Fig.~\ref{fig:exp}. Coupling to both modes is achieved to the electric and magnetic fields of a pin extending from an MMCX connector, also shown in Fig.~\ref{fig:exp}.

Experimental S-parameters are measured using a vector network analyzer connected to the two coupling pins. In practice, ultra-low temperature experiments must use lossy cables to reduce thermal conductivity to the cold stage, and cold attenuators between the source and resonator are often necessary to reduce thermal excitation of microwave photons for experiments ~\cite{huebl}. Attenuation from these make the reflected signal from the resonator weak and difficult to pick out from impedance mismatches in the cabling. Therefore, we only consider the transmission $s_{21}$, shown in Fig.~\ref{fig:s21}, measured at room temperature with a resonator loaded with sapphire wafers. This is compared to calculated transmission data using lossy conductor boundary conditions on the copper walls and sapphire wafers in the gaps, using a mesh size of approximately 2.5~mm, refined further inside and near the gaps where fields vary on a short length scale. The calculated and experimental data qualitatively agree, with quality factors of 510 and 420 measured for the low- and high-frequency modes respectively, with the coupling pins approximately lined up with the sample volume. Disagreement in the shape and mode frequency is likely due to difficulty in modeling the exact pin shape and position, as well as uncertainty in the dielectric constant of sapphire and the magnitude of loss due to oxidation and roughness of the copper surface. This device has been tested down to 0.1~K and in a magnetic field above 5 Tesla. The quality factors increase slightly from room temperature to low temperature, but otherwise the performance of the resonator is not strongly dependent on temperature and it remains mechanically robust and stable in strong magnetic fields as is needed for spectroscopy applications.

\section{Conclusion}

Here we have discussed the finite element analysis of a variation of a typical loop-gap resonator design to provide uniform fields at multiple frequencies. We have shown that this design provides good field uniformity with independent tuning of two modes. Elsewhere, analysis of a five-loop four-gap resonator mentioned the presence of multiple modes within the same volume \cite{eisenach_thesis}, indicating that variations of this basic design could be modified to support more modes with uniform fields. 

Beyond spectroscopic use, these structures could be useful as compact filters with tunability over a range of several GHz.

\end{document}